\documentclass[manuscript,revised]{geophysics}
\usepackage{timet}
\usepackage{natbib}
\setcitestyle{aysep={}}
\usepackage{comment}
\usepackage{xcolor}
\usepackage[colorlinks=true,citecolor=blue,linkcolor=blue,urlcolor=red]{hyperref}
\usepackage{pgfplots}
\pgfplotsset{compat=1.11}
\usepackage{float}
\usepackage[font=small,skip=10pt]{caption}
\usepackage[left= 2.5cm, right=2.5cm, top =2.5cm, bottom=2.5cm]{geometry}
\begin{document}

\title{A feasibility study on monitoring crustal structure variations by direct comparison of surface wave dispersion curves from ambient seismic noise}

\renewcommand{\thefootnote}{\arabic{footnote}}

\address{
\footnotemark[1]Department of Applied Physics, University of Eastern Finland,
Kuopio, Finland}

\author{Kenneth Muhumuza\footnotemark[1]}

\date{}
\maketitle
\date{}
\begin{abstract}
This work assesses the feasibility of the direct use of surface-wave dispersion curves from seismic ambient noise to gain insight into the crustal structure of Bransfield Strait and detect seasonal seismic velocity changes. We cross-correlated four years of vertical component ambient noise data recorded by a seismic array in West Antarctica. To estimate fundamental mode Rayleigh wave Green's functions, the correlations are computed in 4-hr segments, stacked over 1-year time windows and moving windows of 3 months. Rayleigh wave group dispersion curves are then measured on two spectral bands---primary (10--30 s) and secondary (5--10 s) microseisms--using frequency-time analysis. We analyze the temporal evolution of seismic velocity by comparing dispersion curves for the successive annual and 3-month correlation stacks. Our main assumption was that the Green's functions from the cross-correlations, and thus the dispersion curves, remain invariant if the crustal structure remains unchanged. Maximum amplitudes of secondary microseisms were observed during local winter when the Southern Ocean experiences winter storms. The Rayleigh wave group velocity ranges between 2.1 and 3.7 km/s, considering our period range studied. Inter-annual velocity variations are not much evident. We observe a slight velocity decrease in summer and increase in winter, which could be attributed to the pressure melting of ice and an increase in ice mass, respectively. The velocity anomalies observed within the crust and upper mantle structure correlate with the major crustal and upper mantle features known from previous studies in the area. Our results demonstrate that the direct comparison of surface wave dispersion curves extracted from ambient noise might be a useful tool in monitoring crustal structure variations.
\end{abstract}

\section{Introduction}

Passive seismic methods as tools for seismological studies provide excellent opportunities to use continuously recorded seismic noise data, eliminating the need for earthquakes or artificial sources. A widely used method is seismic interferometry, which involves extraction of Green's functions from cross-correlations of ambient seismic noise between station pairs over a sufficient time frame \citep{shapiro2004emergence,sabra2005extracting}. The seismic noise wavefield between a pair of receivers is estimated by creating a virtual source at a receiver location. 

In principle, both surface waves and body waves can be extracted from the cross correlation of noise. However, it has proven to be relatively easy to extract surface waves compared to body waves. This is because most strong seismic noise sources are commonly present within the crust or on the earth's surface \citep{campillo2003long,draganov2009reflection,boschi2015stationary}. Additionally, the majority of ambient noise energy is observed in the microseismic band, which pertains to surface waves of period between  $\sim$5 s and $\sim$30 s \citep{boschi2015stationary}. The microseismic frequency band (5–-30 s) is dominated by surface waves whose frequencies can illuminate crustal structure. This provides the possibility to study earth structure even in areas that lack large seismic networks. Seismic surface waves can be divided into Love waves and Rayleigh waves. Although both of these surface wave types can be observed in the cross-correlations, Rayleigh waves are more preferred since they dominate the vertical component of the seismogram. 

The methodology of retrieving inter-receiver Green's functions from ambient seismic noise data by cross-correlation has led to interesting applications at different scales. Two popular applications include subsurface imaging and monitoring of changes in the subsurface. Subsurface imaging using noise correlations---commonly known as ambient noise tomography---has been achieved using surface waves \citep{shapiro2005high,nicolson2012seismic,obermann20163d} as well as body waves \citep{ryberg2011body,chaput2012imaging,quiros2016seismic}. Monitoring of the processes in the subsurface by chasing changes in seismic waveforms or travel times has been applied successful in the following examples: to monitor geothermal sites \citep{obermann2015potential}, to monitor volcanoes \citep{brenguier2011monitoring,brenguier2008towards}, to monitor fault zones \citep{wegler2007fault,brenguier2008postseismic}, to monitor oil fields \citep{de2013daily}, to monitor landslides \citep{mainsant2012ambient}, to monitor ice sheet melt \cite{mordret2016monitoring} among others.

Monitoring temporal changes in subsurface properties using ambient noise recordings has become very popular. Because noise  continuously illuminates the subsurface, and we can reconstruct waves with virtual sources using only receivers, this technique is useful to detect temporal variations of the earth structure. With this methodology, one obtains some advantages for subsurface monitoring that include repeatability, avoiding the uncertainty in earthquake source locations and origin times, cancelling the complexity of wave propagation to a virtual source, and using measurements (ambient noise) unusable with active source methods. 

In this paper, we use vertical-component of continuous ambient seismic noise data recorded in Antarctica for a 4-year period. We investigate whether Rayleigh-wave group dispersion curves based on ambient noise surface wave data can be compared directly to monitor seasonal variations in seismic velocities. This is on the assumption that if there are no changes, noise measurements performed on the same earth medium at different seasons yield the same dispersion characteristics. Our analysis is concentrated on two different period bands: 5–-10 and 10-–30 s. We explore in detail the cross-correlation results for the JUBA--ESPERANZA pair, with a path length of approximately 154 km connecting King George Island and the Antarctic Peninsula. We examine the implications of seasonal variations on Southern Ocean microseisms. We obtain insight on the feasibility of the direct use of Rayleigh wave dispersion curves to study the crustal structure of the Bransfield Strait and monitor seasonal velocity variations.

\section{Data and measurement methods} 

\begin{figure}[H]
\begin{center}
\includegraphics[width=0.8\textwidth]{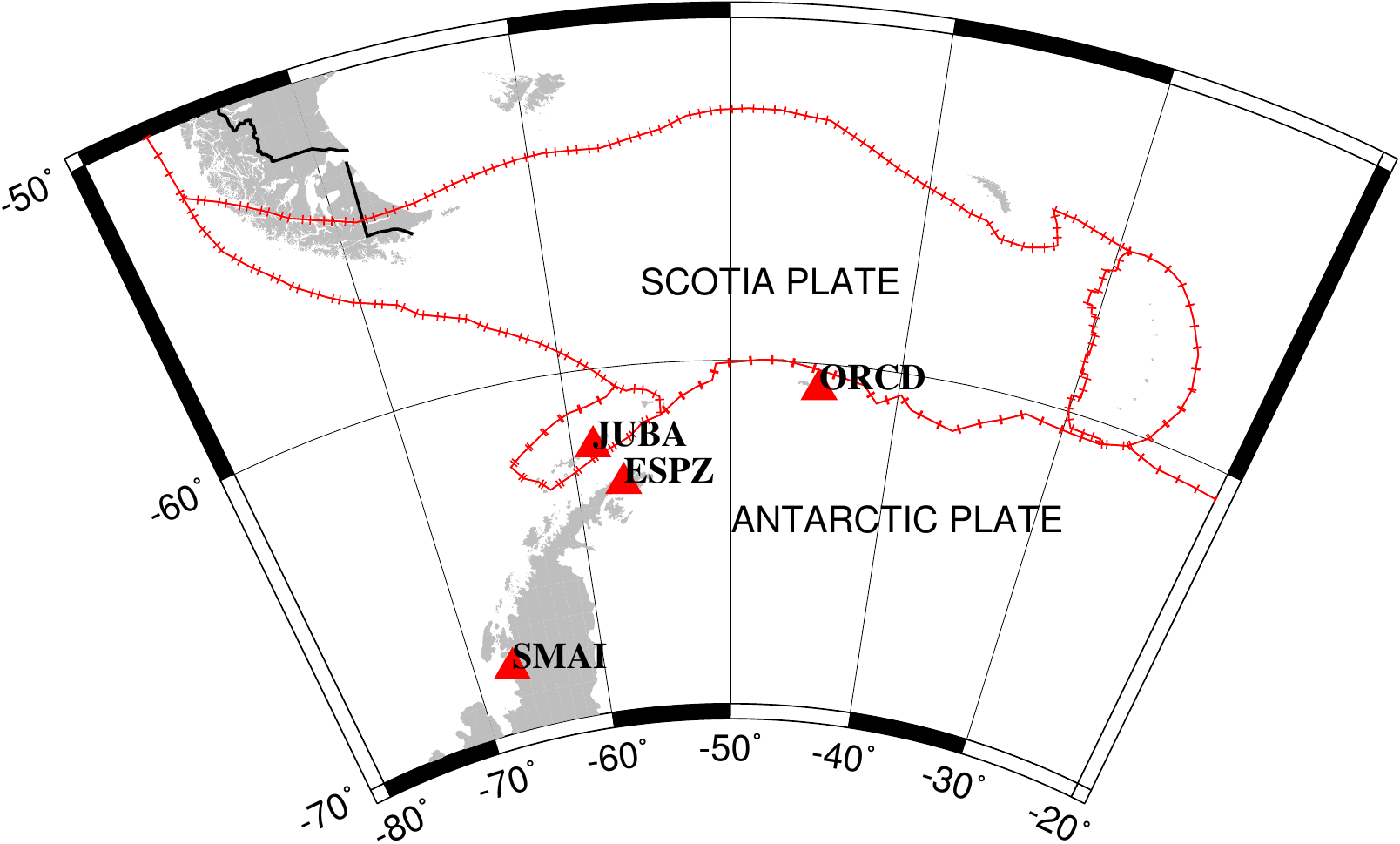}
\end{center}
\caption{Location of seismological stations within the ASAIN network. Red triangles labelled JUBA, ESPZ,   ORCD and SMAI are stations used in this study. The red lines indicate plate boundaries.}
  \label{figure1}
\end{figure}

In this study, we used vertical component continuous seismic data from 4 stations in the Argentine Antarctica region from ASAIN (Antarctic Seismographic Argentine-Italian Network) in West Antarctica (Fig.~\ref{figure1}). The stations Carlini---formerly Jubany (JUBA)---station, Esperanza (ESPZ), Orcada (ORCD) and San Martin (SMAI) are part of the 5 seismological stations that have been set under ASAIN collaborative agreement. The seismologic network  provide essential data to study the internal structure of the Antarctic continent, monitor seismicity, and to monitor large-scale phenomena, such as  melting of the Antarctic ice sheet, among others. We analyzed 4 years (2008-2011) of seismic data recorded recorded by the 4 stations with a sampling rate of 1 sample per second. The initial data preparation involves unpacking the raw data from its standard SEED format to obtain noise data in SAC format. We have to ensure that the two amplitudes of each record are completely consistent by running a time normalization so that the records share the same absolute time and the same amplitude information is retained. It is important to note that if the instruments of two stations are inconsistent, it is necessary to remove instrument responses from the raw data.

Here we concentrate our feasibility analysis on the JUBA-ESPZ pair whose path crosses the Bransfield Strait from the King George Island to the Antarctic Peninsula. The goal of our study is to gain insight into the crustal structure of the Bransfield Basin and detect seasonal seismic velocity changes. This basin has been studied by various authors for over the last five decades \citep{ashcroft1972crustal,janik1997seismic,grad1997crustal,christeson2003deep,barker2003backarc,vuan2005crustal,janik2006moho,yegorova2011joint,park2012p,vuan2014reappraisal}. But the details of its crustal structure has been difficult to describe and the conclusions have been controversial. 

\section{Techniques for estimating temporal seismic velocity variations}

The use of noise correlations as repeated seismic sources allows us to compare waveforms at different time intervals. This implies that temporal variations in the earth structure (measured as velocity variations) can be monitored by means of subtle differential characteristics of the reference and current signals. The reference is usually obtained by stacking daily cross-correlations over a long period of time and the current by stacking over a few days. For applications that involve temporal monitoring, it is assumed that the seismic Green's function from the cross-correlation of ambient seismic noise remain unchanged if the investigated medium is stationery. Efficient preprocessing of noise to correct for seasonal variation is important else variation of noise sources could result into changes that may be mistakenly interpreted as medium temporal variations \citep{hadziioannou2009stability,weaver2011precision,gong2015effects}. 

There are two different methods that have widely been applied to study temporal changes in different earth structures by measuring velocity perturbations $dv/v$. They include the "Stretching" Technique (ST) \citep{sens2006passive,hadziioannou2009stability} and the Moving Window Cross Spectral technique (MWCST) \citep{clarke2011assessment}. The important consideration in both methods is that changes in the properties of the medium in which the seismic waves propagate result in seismic waveforms or travel times changes. Thus, relative velocity changes in the medium are estimated by comparing the two waveforms: reference and current cross-correlation signal waveforms.

The ST operates in the time domain and is based on the assumption that if a homogeneous velocity perturbation occurs in the medium, then travel times of the seismic phases change in proportion to the perturbation. This results in a stretched or compressed current correlation waveform in comparison to the reference \citep{meier2010detecting}. The stretching parameter $\epsilon _{m}$ that maximises the correlation coefficient of the two waveforms in a given lapse time window $t$ is such that, 
\begin{equation}
\epsilon _{m}=-dt/t=dv/v. 
\label{eq3}
\end{equation}
 
The MWCST involves for a given time window, computing the cross-spectrum  between the current and the reference correlation function to estimate time delays. The time delay measurements for different time windows are estimated from the slopes of the phase of the cross-spectra. Then, the final value of $dt/t$ is obtained using a linear regression model \citep{duputel2009real}. The $dv/v$ value in the medium is deduced by a similar relation in Equation~\ref{eq3}. Since the MWCST operates in the frequency domain, it may be less affected by seasonal variations in noise sources than the ST that operates directly in the time domain.  

For monitoring purposes involving tiny medium changes, the two methods above are preferably applied to the coda part of the cross-correlation function that appears after the first (direct) arrival. This is because coda waves  sample the medium repeatedly and thus more sensitive to changes in the medium properties. However, we believe that for significant changes in the medium, analysis of the time window of the direct arrival Rayleigh waves in the the cross correlation should be sufficient. Moreover, compared to the coda, the direct surface waves are sensitive to structures in the top few kilometres between a station pair; hence they are suitable for studying the earth's crust and detecting temporal variations in crustal properties. In this paper, we measure group velocity dispersion curves for different temporal ranges to detect possible seasonal seismic velocity changes on a selected path across the Bransfield Strait in West Antarctica. More complete statistics on dispersion curves can also be useful in estimating the depth of temporal changes by utilising depth sensitivity of Rayleigh waves.

\section{Estimation of Green's function}
We estimate Rayleigh-wave Green's functions by computing the cross-correlations of ambient-noise records between a pair of seismologic stations. Before the actual cross correlation is made, a noise preprocessing step is as follows: each daily trace is cut into 4-hour long segment of noise, and each resulting segment is detrended, demeaned, tapered, and band-pass filtered between periods of 1 and 100 s. To eliminate the relevance of spurious spikes and reduce effects of earthquakes on cross correlations, we activate one-bit normalisation \citep{bensen2007processing}, a method presently widely used. As a final preprocessing step, we apply spectral whitening to the data. 

For a station pair, the 4-hour correlations were linearly stacked into moving averages of three months, that is to say, for each year we consider 12 overlapping 3-month time windows starting from January; namely January, February, March; February, March, April; ... November, December, January (next year). We find the 3-month correlation stacks suitable for determination of group velocity dispersion curves because they produce inter-station wavefields with acceptable signal-to-noise ratios. Therefore, we can analyze four different temporal intervals for every year to estimate variations in group velocity measurements that describe temporal variations in crustal structure. 

It has been shown that sufficient spectral whitening of the cross-correlated signals can potentially remove the effects on the ambient noise cross-correlations of the distribution of ambient noise sources if the seasonal variations are uniform with respect to the noise source locations \citep{daskalakis2016robust}. This hypothesis is found reasonable when the measurements are from the same regional area. We have performed an adequate normalization, but still found small differences in the frequency domain amplitude spectra at low periods, depending on the kind of path.

To ensure the reliability of our measurements in each period band, we considered cross-correlations whose SNR $>10$. The SNR was computed as the ratio of the maximum amplitude of the correlation within a time window containing the signals (2- to 4-km/s window) to the root-mean-square of noise slower than the signal arrival window (slower than 2 km/s). Hence, the estimated Green's functions are the computed cross-correlations with SNR $> 10$. For the JUBA-ESP pair studied in detail here, the SNR of the correlations is the highest and the correlations within the analyzed frequency band are least biased by seasonal variations of the noise. In general, however, a robust kind of correction for seasonal variations would be necessary especially when working with low periods. 

\section{Seasonal variations of microseismic signal}

\begin{figure}[H]
  \centering
  \includegraphics[width = \textwidth]{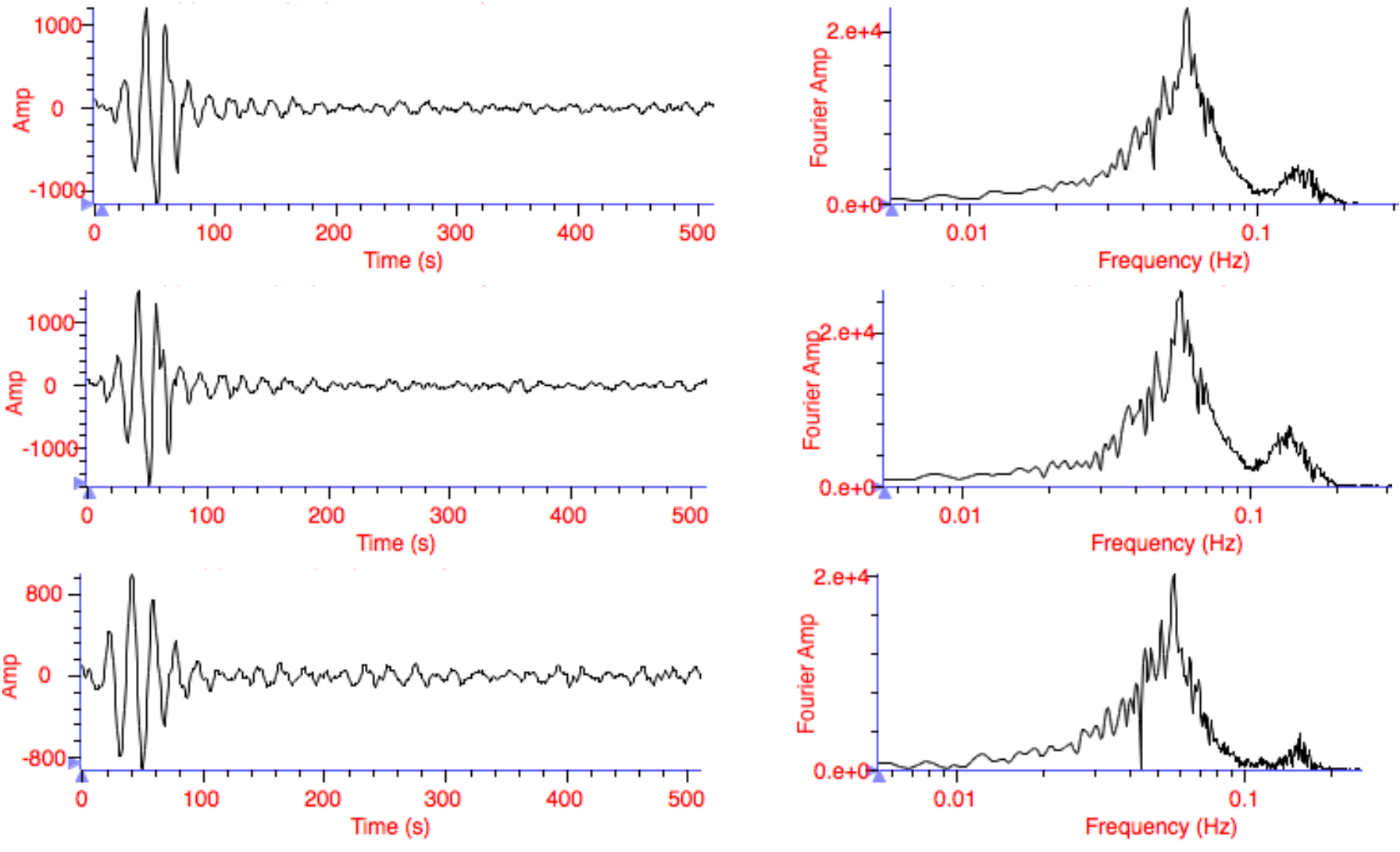}
\caption{Cross-correlations of three months of noise data (January to March 2010) between ESPZ and JUBA, with a path length of approximately 154 km. The amplitude spectrum is also shown to the right of each waveform. Upper panel: Symmetric-component of the cross-correlation (folded signal). Middle panel: The negative part of the cross-correlation. Lower panel: The positive part of the cross-correlation. The negative and positive parts of the cross-correlation represent two opposite directions of wave propagation between two receivers.}
  \label{figure2}
\end{figure}

Seasonal variations of primary and secondary microseisms can be observed in the empirical Green’s function retrieved from the cross-correlation traces. Fig.~\ref{figure2} illustrates the asymmetry of cross-correlations in our data. For example the middle panel of Fig.~\ref{figure2} shows that the negative lag of the cross correlation is predominant in the secondary microseism band  while the positive lag component in the lower panel is weak. This indicates that the ambient noise travelled dominantly in a direction from JUBA towards ESPZ, so generally from a station near the coast since the ocean is the dominant noise source. Without clear knowledge about which component of the signal is better,  the folded signal (upper panel of Fig.~\ref{figure2}), which results from time reversing the negative lag and summing it with the positive one, is a better signal. By averaging the positive and negative components, the signal to noise ratio of the empirical Green's functions is improved \citep{buffoni2018rayleigh}. 

In this study, we use the symmetrical component (folded signal) to extract the dispersion curves. As earlier pointed out, there could be a high possibility of effects caused by the variation of noise sources around Antarctica affecting the shape of negative and positive parts of the cross-correlation. Since noise sources for the secondary microseisms appear relatively stable, it is suitable to focus on the frequency range of secondary microseisms when analyzing the dispersion curves.

\begin{figure}[H]
\centering
\includegraphics[width = \textwidth]{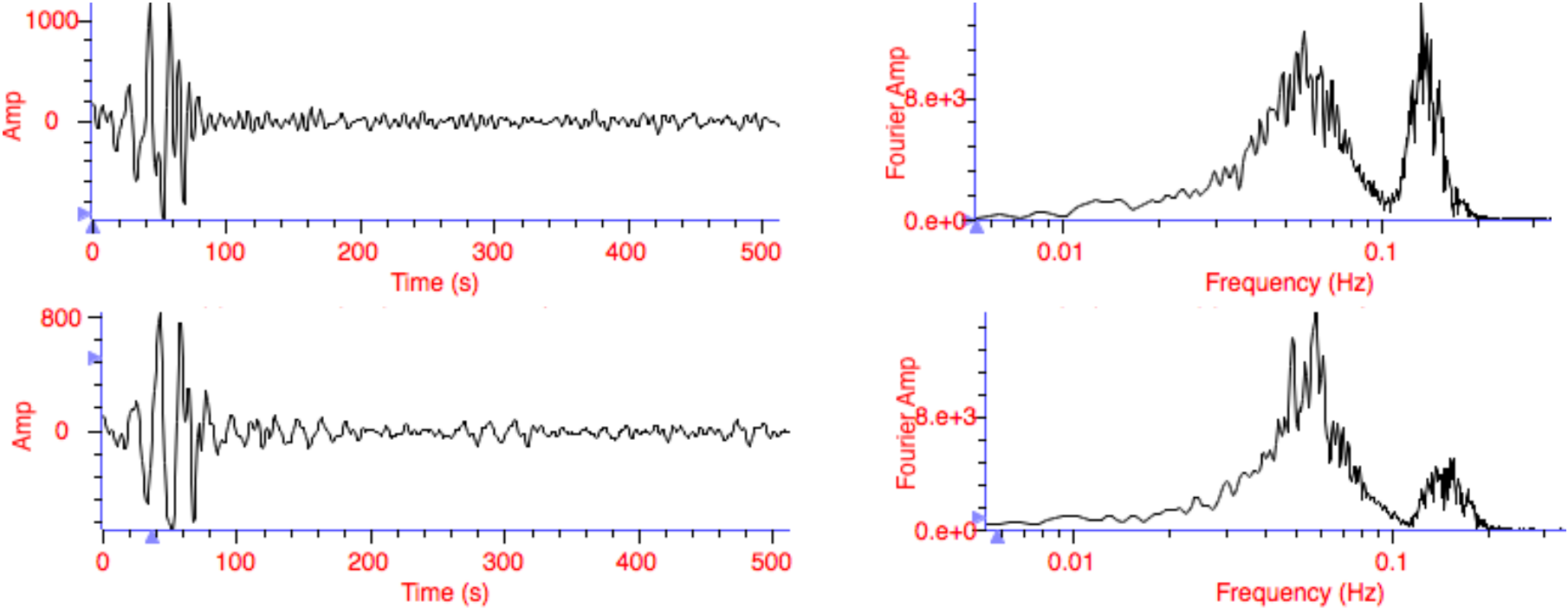}
\caption{Cross-correlations of noise data between ESPZ and JUBA for three months of Antarctica local summer and local winter season in 2008. The amplitude spectra are also shown to the right of each waveform. Upper panel: Folded cross correlation of three months of Antarctica local winter (June to August). Lower panel: Folded cross correlation of three months of Antarctica local summer (January to March).}
\label{figure3}
\end{figure}

Fig.~\ref{figure3} shows, for example, the estimated Rayleigh-wave Green's functions between station ESP and JUBA for temporal ranges January--March and June--August in 2008. The amplitude spectra of the signal is also shown on the right side. The period band of 3--30 s is the most energetic. Two predominant peaks in a frequency domain corresponding to primary (10--30 s) and secondary (5--10 s) microseisms can be clearly seen. Although these seismic records are believed to originate from long period ocean waves, the secondary microseisms are generated by a more complex mechanism than the primary microseisms.

Fig.~\ref{figure3} also shows that typical noise correlation signals vary over time. We observe that the secondary microseism noise level is minimum during local summer, with higher levels occurring during local winter when the Southern Ocean experiences violent winter storms. The relatively higher secondary microseisms in austral winter could also be explained by instances of ice‐breakups and near-coastal interactions related to iceberg reflections \citep{grob2011observations,davy2016analyses,pratt2017implications,lepore2018analysis}. For the primary microseism band, noise levels are reduced in the austral winter. As sea ice builds out along the coastline during the winter season, its eliminates ocean wave forcing on the Antarctic continental shelf \citep{grob2011observations,pratt2017implications}. The presence of sea ice therefore potentially decreases the generation of primary microseisms, leading to relatively low primary microseism spectral amplitudes. The increase in the wave shoaling rates after the loss of sea ice may explain the much stronger amplitudes of the primary microseisms during austral summer. Because the primary microseisms are largely generated at the coast and highly dependant on sea ice at near-coastal locations, they could strongly be affected by seasonal variations and the noise sources. 

\section{Group velocity dispersion measurements}
\begin{figure}[H]
\centering
\includegraphics[width = \textwidth]{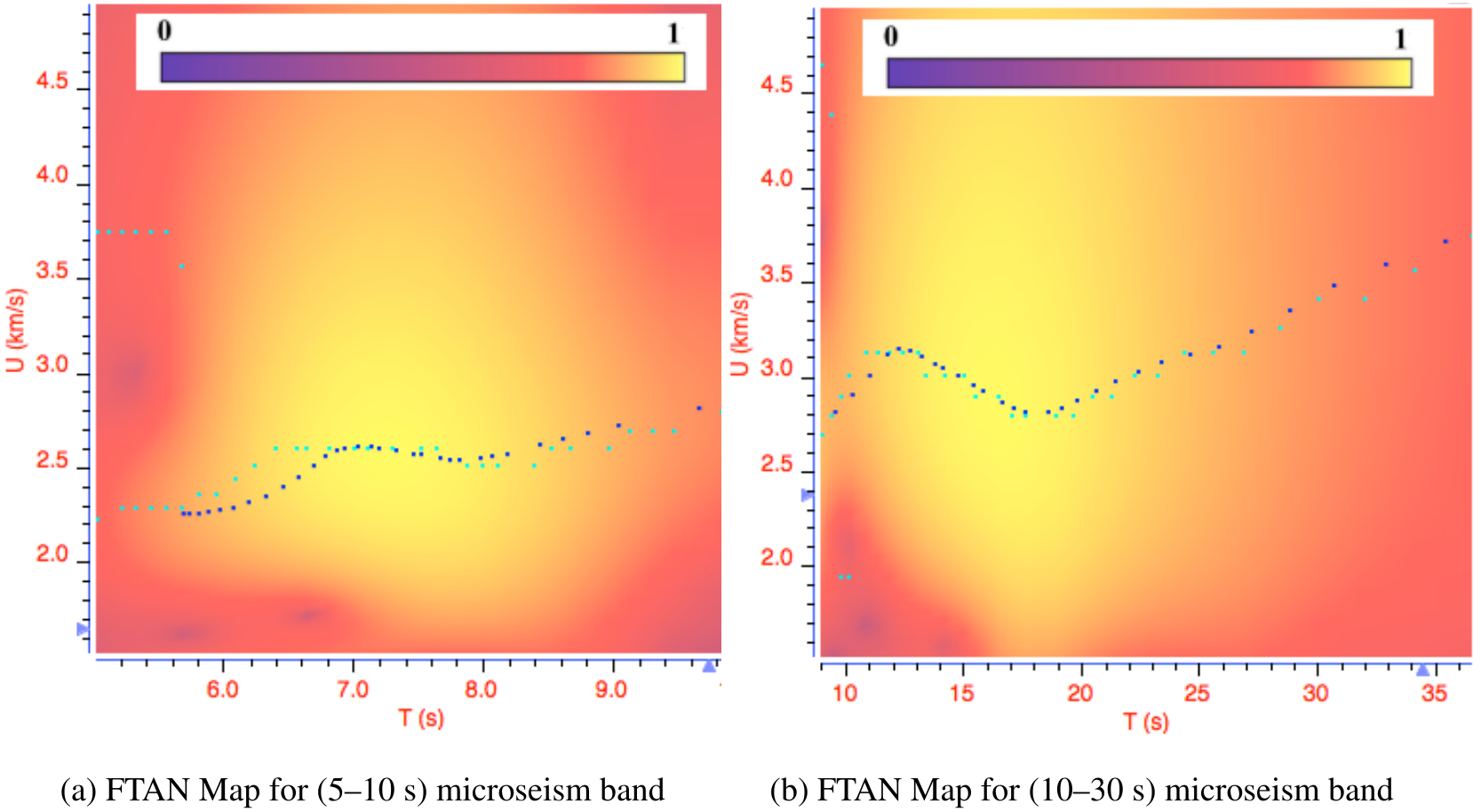}
\caption{Example of FTAN map showing Rayleigh wave group velocity for the path ESPZ-JUBA for the spectral band corresponding to two frequency bands: (a) secondary (5--10 s) microseism band (b) primary (10--30 s) microseism band. The yellow areas in the map are those associated with the energy arrivals. Cyan dots indicate relative maxima for a given filter central period. Blue dots indicate the period corresponding to the fundamental mode of Rayleigh waves. In this figure, we show the overall energy distribution normalized to 1, whereas the energy was normalized for each frequency to improve the tracking of group arrivals. The measured group velocity values (U) are plotted against the filtered central period (T)}. 
\label{figure4}
\end{figure}

We measure group velocity  dispersion curves on the estimated Green's function of each temporal range by applying frequency time analysis (FTAN), which gives for each period, an estimate of the surface wave group velocity \citep{dziewonski1969technique,levshin1989seismic,bensen2007processing}. Dispersion curves can be visualized using FTAN map. Figure~\ref{figure4} illustrates a typical dispersion diagram obtained for the station pair ESPZ--JUBA. A line joining the blue dots indicates the fundamental mode of the Rayleigh waves. The fundamental mode of the Rayleigh waves at periods greater than 5 s is clearly evident and can actually be identified on the station pairs for the different temporal ranges. 

The dispersion curves in our study show a marked change at periods around 5 s, and it is hard to estimate the group velocities at periods shorter than 5 s. This trend may be as a result of the presence of ice strongly that attenuates the short period seismic noise or the small inter-station distance in our case. Reliable dispersion measurements at longer periods are  restricted by inter-station distances since stations must be separated atleast three wavelength \citep{bensen2007processing}. In this study, with inter-station paths ranging from 150--800 km, we can focus our analysis in the period range 2--30 s.  We obtain good dispersion curves with strong frequency content between 6.0 seconds and up to 30 seconds.

The dispersion map is represented by a complex function that includes a narrow band pass filter. The optimal choice used in the FTAN technique is a Gaussian filter which is described by $exp\left [ -\alpha(f-f_{c})^{2}/f_{c}^{2}  \right ]$. The $\alpha$ value determines the the frequency resolution of the filter and $f_{c}$ is the central frequency. The parameter $\alpha$ defines the width of the band-pass filter. This parameter must be properly chosen to have a meaningful shape of the dispersion curve and achieve a good balance between time and frequency resolution. We have explored a wide range of values for $\alpha$ and found that the best values for our study are between 20.0 and 60.0, depending on the inter-station distance and period range studied. From Fig.~\ref{figure4}, its clear than the two spectral bands corresponding to the primary (10--30 s) and secondary (5--10 s) microseism can be analysed separately to have more meaningful and accurate dispersion curves. An appropriate value of $\alpha$ for each period range must be selected during the analysis. 

The measured group velocity dispersion curves can potentially be affected by seasonal variations in the noise sources. To validate that our observed group velocity variations may be due to crustal changes, we estimate uncertainties for the group velocities based on seasonal variations of the dispersion curves. The error analysis procedure involves using twelve overlapping 3-month correlation stacks to investigate the seasonal variability of the measurement. The 3-month time windows not only provide reliable dispersion measurements, but also mainly capture seasonal fluctuations of the noise sources.

For each year of data and period range, we computed the standard deviation on all the overlapping 3-month correlation stacks. We find that our group velocity uncertainties are smaller in the 5--10 s period range and tends to increase with period. Larger uncertainties are prevalent at periods larger that 20 s due to our shorter path length and perhaps because the amplitude of ambient noise decreases at periods greater than 20 s \citep{yang2007ambient}. Fig~\ref{figure7}, for example, shows group velocity measurements in twelve three-month cross-correlations for the 3-months moving windows. The one-year reference is plotted as a black line with error bars corresponding to the computed standard deviation.

While some of the variability we are seeing may be due to variation in the noise sources, the results indicate that the noise preprocessing is efficient in mitigating noise source variations. Hence most of the group velocity variations we observe in the 5--20 s period range is mainly associated with changes in the seismic velocity structure rather than changes in the noise field. Moreover, in this frequency range, the seismic waves mostly sample the Antarctica crust between depths of 5 and 30 km.

\section{Velocity variations}
We assess velocity variations for the JUBA-ESP path by dispersion curve analysis. The assumption is that changes in the crustal structure should result into changes in velocities and hence the measured dispersion curves. Fig.~\ref{figure6} shows yearly variations of Rayleigh wave group velocity corresponding to four years (2008--2011). It can be immediately seen that the group velocity values range within 2.1-–3.7 km/s, considering that the group dispersion curves have been calculated for a 5--30 s period range. For a suitable value of $\alpha$ (Figs~\ref{figure6}a and \ref{figure6}b), we observe systematic variations for all the years with very small changes in the group velocity of the order of less than $\pm 0.05\%$. Based on this observation, we could suggest that inter-annual velocity change  along the JUBA-ESP path in 2008--2011 are not evident. Hence, for a year average, the effects of uneven and varied spatial distribution of noise sources do not affect the measured dispersion curves. This is because the effects cancel out in the comparison. However, they may affect the accuracy of the actual velocities obtained and not the velocity variations. Fig.~\ref{figure6}d shows the effect of a high $\alpha$ parameter in estimating velocity variations. The curves give misleading and less systematic velocity variations when there is no balance between time and frequency resolution. 

To monitor seasonal velocity changes and study the effect of noise source variation, we compare the measured dispersion curves for 3-month moving windows. Figs~\ref{figure7}, (and Figures S1, S2 and S3 in Supplementary Material) show the measured temporal variations of dispersion curves (Rayleigh wave group velocities) for the years 2008--2011, respectively. Superimposed is the dispersion curve (black line) for a stack of each year, which acts as a good reference. The temporal group velocities measured are reasonable, varying around our reference Rayleigh wave group velocity dispersion curve. We observe a slight velocity decrease in summer, which could be attributed to the pressure melting of ice, similar to what has been reported for station pairs in the coastal areas \citep{toyokuni2018changes}. It has also been suggested that an increase in ice mass during winter months could result in velocity increase due to the compaction of ice and bedrock, which changes the stress field. However, by comparing with the reference, the temporal changes in dispersion curves could also be influenced by the seasonal distribution of noise sources, resulting in seasonal variations of the amplitude spectra of ambient noise. It is observed that this temporal variability mostly affects the measurements for periods $T > 10$ s. Moreover, these less systematic variations appear in the period ranges where the amplitude spectra peaks. Hence, when using dispersion curves derived from ambient noise correlations as a tool to measure velocity variations, correcting for seasonal variations is important. 

\begin{figure}[H]
    \centering
    \includegraphics[width=0.5\textwidth]{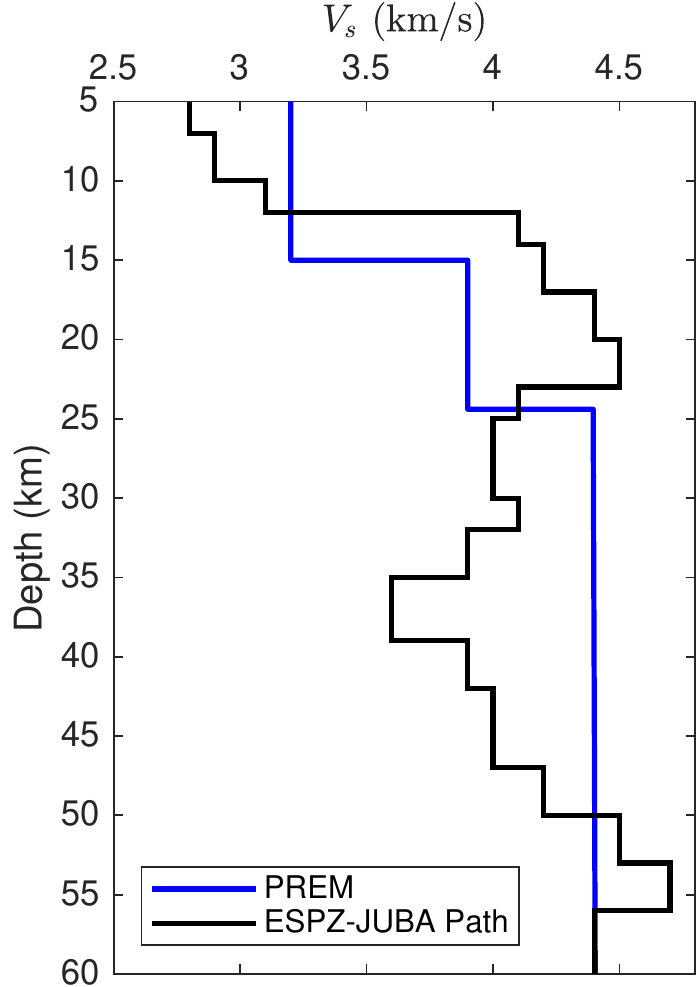}
\caption{Shear velocity-depth profile for the ESPZ-JUBA path (black line). The starting $V_{s}$ model, which is the Preliminary Reference Earth Model (PREM) $S_{V}$ global model is shown in blue.}
    \label{figure5}
\end{figure}

\begin{figure}[H]
\centering
\includegraphics[width=0.95\textwidth]{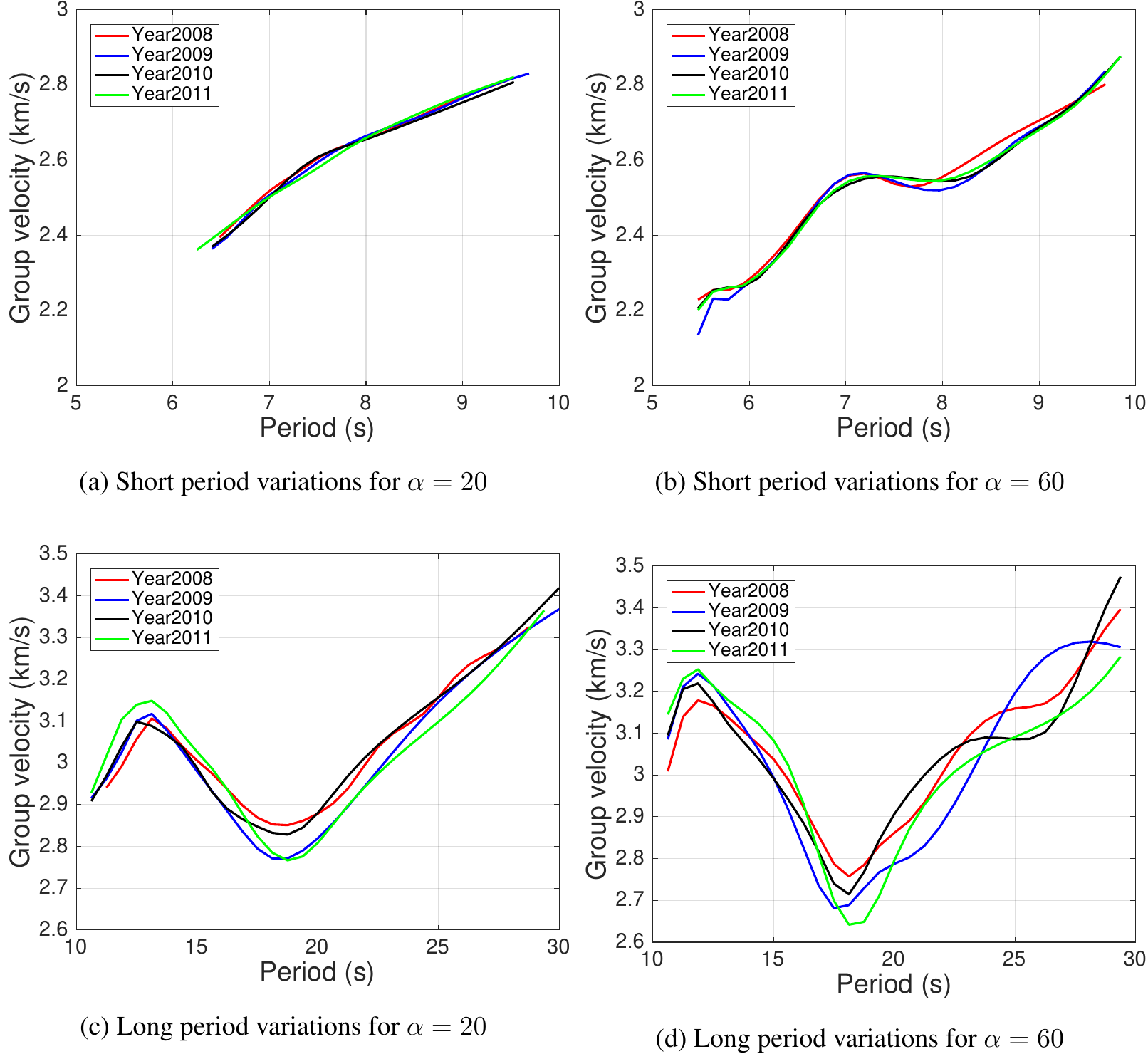}
  \caption{Variation in the yearly measured  dispersion curves for the time period 2008--2011. Upper panel: Short period (5--10 s) band. Lower panel: Long period (10--30 s) band. Each line represents a dispersion curve for a year, color coded by the year (see legend).}
  \label{figure6} 
\end{figure}

From our measured Rayleigh wave group dispersion curves, we used the code SURF96 \citep{herrmann2013computer} and a starting model PREM \citep{dziewonski1981preliminary} to estimate a velocity-depth profile for ESPZ-JUBA path. The other input parameters for the inversion were based on the starting model. It has been indicated earlier that the PREM shows a velocity structure that is similar to that of West Antarctica \citep{heeszel2016upper}. The 1-D $V_{s}$ model for the ESPZ--JUBA path obtained from our data-set is shown in Fig~\ref{figure5}. A high velocity zone is observed at depth intervals of about 14--19 km up to 25km, as well as 50--55 km. Previous studies in the central part of Bransfield Strait \citep{janik1997seismic,grad1997crustal,janik2006moho} have revealed similar observations of a high velocity body that extends from a depth of 13 km down to the Moho boundary that reaches a depth of 42 km.  A low velocity zone, which underlies the high velocity zone is also detected at 25--30 km, 32–-40 km depth, and as well as below 55 km. About 10\% decrease in group wave velocity is observed between depths 25 and 40 km. Previous studies have suggested the existence of low-velocity anomalies in the lower crust-uppermost mantle structure of Bransfield Strait \citep{christeson2003deep,vuan2005crustal,yegorova2011joint,park2012p,vuan2014reappraisal}, which are attributed to a thermal anomaly in the upper mantle, active volcanism, and the rifting of the continent. 

For longer periods ($>$20 s), we obtain inaccurate velocity variations as shown for example in Figs~\ref{figure7}f, and (Figure S1(f) in Supplementary Material). This implies that the condition for stations to be separated at least three wavelengths to have reliable dispersion measurements is important even for monitoring using dispersion curves. For instance in our case, with a typical surface wave travelling at 3 km/s and a path of 154 km, the condition requires that only we analyze up to about a period of 20 s. In this period range, Rayleigh waves are sensitive to velocities in the crustal thickness and uppermost mantle up to about 55 km for a typical ocean basin \citep{ritzwoller2001crustal}.

\begin{figure} [H]
\centering
\includegraphics[width = 0.9\textwidth]{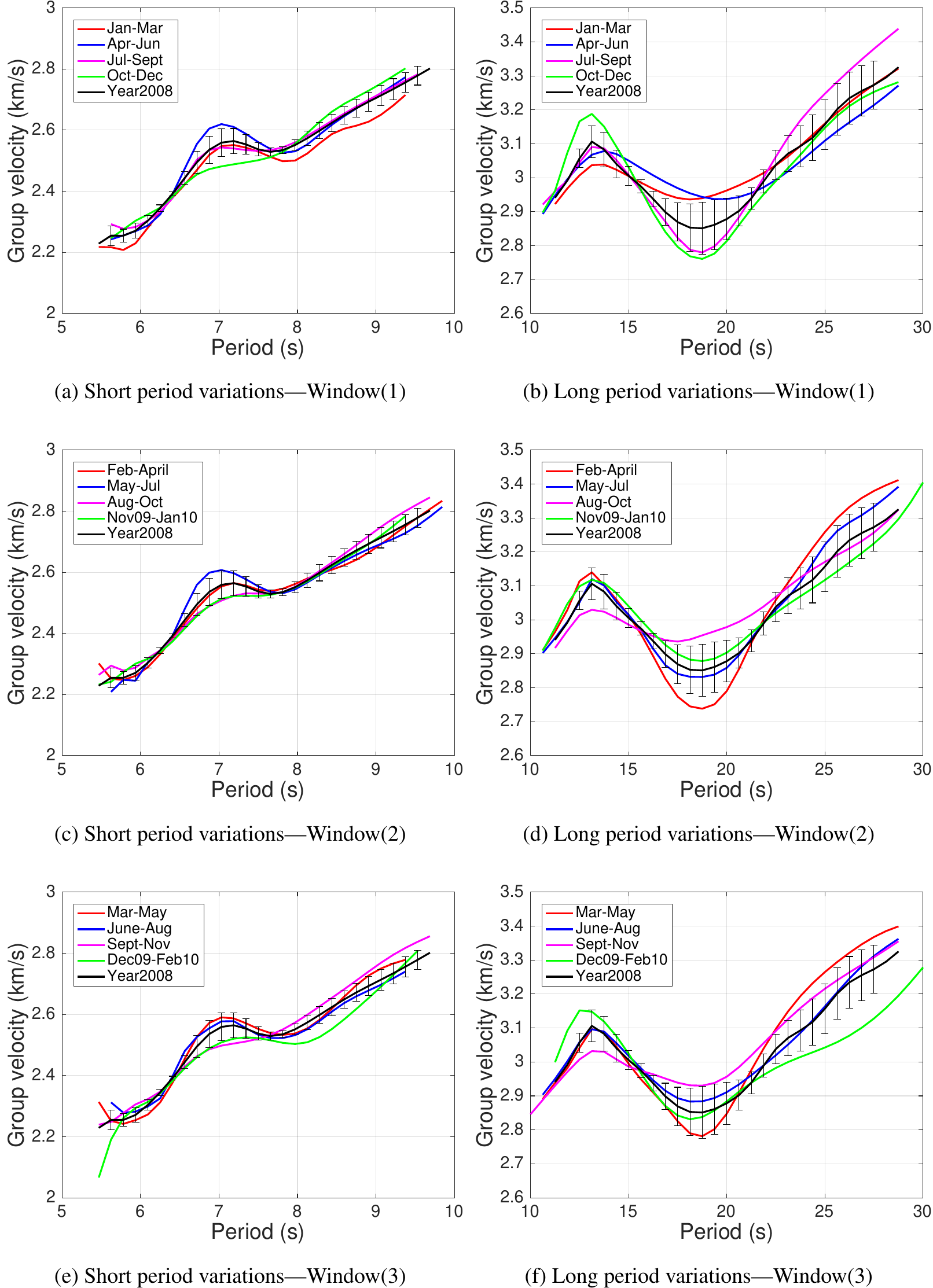}
  \caption{Variation in the measured  dispersion curves for 3-month range moving windows of 2008 for short period (5--10 s) and long period (10--30 s) bands. Black line is the one year reference, and the error bars shown correspond to the computed standard deviation.}
  \label{figure7} 
\end{figure}

\section{Conclusions}

 We assess the feasibility of monitoring seasonal velocity variations by direct comparison of Rayleigh wave group velocity dispersion curves extracted from ambient noise correlations. We present results for a selected path that crosses the Bransfield Strait almost perpendicular to its main axis. Using continuous ambient seismic noise recordings for a 4-year period, we cross-correlate 4-hr long segments, stack the cross-correlations over 3-month moving windows, and perform analysis for every quarter. In the microseismic period band (5-30 s) studied, we observe seasonal variations of vertical component of seismic noise spectra. Secondary microseism noise level is maximum during local winter when the Southern ocean experiences winter storms. The seasonal fluctuations of the amplitude spectra of the noise sources may affect our interpretation of the temporal velocity changes and thus their effect should be corrected. The 5 s–-30 s group velocity dispersion curves show low- and high-velocity anomalies that are mostly correlated with crust and upper mantle structure. At periods $<$5 s, it is hard to estimate the group velocities, perhaps because of the the small inter-station distance and the fact that the presence of ice strongly attenuates the short period seismic noise. Our results corroborate with many of the previous ones obtained on profiles along the paths connecting the Shetland Islands and Antarctic Peninsula. We observe a slight velocity decrease in summer and increase in winter, which could be associated with both broad-scale and local sea-ice conditions. Compared to the inter-seasonal (summer to winter) surface velocity variations, there were no significant inter-annual variations observed. This observation could suggest that the cause for inter-seasonal changes is mostly localized while the annual variations are linked to broad-scale conditions. The results demonstrate that one may directly compare Rayleigh-wave group velocity as useful tool to monitor seasonal changes in the crustal conditions.
 
 A detailed interpretation of Rayleigh-wave group dispersion curves based on a denser seismic network is part of our future plans. This would help to test whether dispersion curves for large inter-station distances give meaningful results in the temporal variation estimation process. In our suggested approach, the monitoring of seismic velocity changes is achieved without inversion of Rayleigh wave dispersion curves for 2D shear wave velocity profiles. We believe that forward modelling would provide the possibility to use a suitable structure and compare theoretical curves obtained in order to see how the variations affect the dispersion curves. By modelling seismic velocity variations, one can suggest mechanisms for ice mass loss variations locally from seismic noise measurements, though this is not part of this work.

\segsection*{acknowledgments}
This paper is based on the author's Postgraduate Diploma thesis completed at the Abdus Salam International Centre for Theoretical Physics (ICTP), Trieste. The Computational Seismology Group at the department of Mathematics and Geosciences, University of Trieste provided the data and softwares used in this study. I wish to extend my sincere appreciation to Professor Romanelli Fabio, Davide Bisignano, and Franco Vaccari for valuable suggestions and technical support throughout the development of this work. The author gratefully acknowledge the financial support from the University of Eastern Finland through the Finnish Centre of Excellence of Inverse Modelling and Imaging. I would like to thank the editor and two anonymous reviewers for their valuable comments and input that helped to improve this article.

\segsection*{Supplementary Material}
Supplementary material for this article includes:
\begin{itemize}
    \item Figure S1. Variation in the measured dispersion curves for 2009.
    \item Figure S2. Variation in the measured dispersion curves for 2010.
    \item Figure S3. Variation in the measured dispersion curves for 2011.
\end{itemize}

\bibliographystyle{seg}  
\bibliography{refs}

\begin{thebibliography}{51}
\expandafter\ifx\csname natexlab\endcsname\relax\def\natexlab#1{#1}\fi

\bibitem[Ashcroft(1972)]{ashcroft1972crustal}
Ashcroft, W.~A., 1972.
\newblock {\it Crustal structure of the South Shetland Islands and Bransfield
  strait\/}, vol.~66, British Antarctic Survey.

\bibitem[Barker et~al.(2003)Barker, Christeson, Austin, \&
  Dalziel]{barker2003backarc}
Barker, D.~H., Christeson, G.~L., Austin, J.~A., \& Dalziel, I.~W., 2003.
\newblock Backarc basin evolution and cordilleran orogenesis: insights from new
  ocean-bottom seismograph refraction profiling in bransfield strait,
  antarctica, {\it Geology\/}, {\bf 31}(2), 107--110.

\bibitem[Bensen et~al.(2007)Bensen, Ritzwoller, Barmin, Levshin, Lin,
  Moschetti, Shapiro, \& Yang]{bensen2007processing}
Bensen, G., Ritzwoller, M., Barmin, M., Levshin, A., Lin, F., Moschetti, M.,
  Shapiro, N., \& Yang, Y., 2007.
\newblock Processing seismic ambient noise data to obtain reliable broad-band
  surface wave dispersion measurements, {\it Geophysical Journal
  International\/}, {\bf \textbf{169}}(3), 1239--1260.

\bibitem[Boschi \& Weemstra(2015)]{boschi2015stationary}
Boschi, L. \& Weemstra, C., 2015.
\newblock Stationary-phase integrals in the cross correlation of ambient noise,
  {\it Reviews of Geophysics\/}, {\bf \textbf{53}}(2), 411--451.

\bibitem[Brenguier et~al.(2008{\natexlab{a}})Brenguier, Campillo, Hadziioannou,
  Shapiro, Nadeau, \& Larose]{brenguier2008postseismic}
Brenguier, F., Campillo, M., Hadziioannou, C., Shapiro, N., Nadeau, R.~M., \&
  Larose, E., 2008{\natexlab{a}}.
\newblock Postseismic relaxation along the san andreas fault at parkfield from
  continuous seismological observations, {\it science\/}, {\bf
  \textbf{321}}(5895), 1478--1481.

\bibitem[Brenguier et~al.(2008{\natexlab{b}})Brenguier, Shapiro, Campillo,
  Ferrazzini, Duputel, Coutant, \& Nercessian]{brenguier2008towards}
Brenguier, F., Shapiro, N.~M., Campillo, M., Ferrazzini, V., Duputel, Z.,
  Coutant, O., \& Nercessian, A., 2008{\natexlab{b}}.
\newblock Towards forecasting volcanic eruptions using seismic noise, {\it
  Nature Geoscience\/}, {\bf \textbf{1}}(2), 126--130.

\bibitem[Brenguier et~al.(2011)Brenguier, Clarke, Aoki, Shapiro, Campillo, \&
  Ferrazzini]{brenguier2011monitoring}
Brenguier, F., Clarke, D., Aoki, Y., Shapiro, N.~M., Campillo, M., \&
  Ferrazzini, V., 2011.
\newblock Monitoring volcanoes using seismic noise correlations, {\it Comptes
  Rendus Geoscience\/}, {\bf \textbf{343}}(8), 633--638.

\bibitem[Buffoni et~al.(2018)Buffoni, Schimmel, Sabbione, Rosa, \&
  Connon]{buffoni2018rayleigh}
Buffoni, C., Schimmel, M., Sabbione, N.~C., Rosa, M.~L., \& Connon, G., 2018.
\newblock Rayleigh waves from correlation of seismic noise in great island of
  tierra del fuego, argentina: constraints on upper crustal structure, {\it
  Geodesy and Geodynamics\/}, {\bf 9}(1), 2--12.

\bibitem[Campillo \& Paul(2003)]{campillo2003long}
Campillo, M. \& Paul, A., 2003.
\newblock Long-range correlations in the diffuse seismic coda, {\it Science\/},
  {\bf \textbf{299}}(5606), 547--549.

\bibitem[Chaput et~al.(2012)Chaput, Zandomeneghi, Aster, Knox, \&
  Kyle]{chaput2012imaging}
Chaput, J., Zandomeneghi, D., Aster, R., Knox, H., \& Kyle, P., 2012.
\newblock Imaging of erebus volcano using body wave seismic interferometry of
  strombolian eruption coda, {\it Geophysical Research Letters\/}, {\bf
  \textbf{39}}(7).

\bibitem[Christeson et~al.(2003)Christeson, Barker, Austin, \&
  Dalziel]{christeson2003deep}
Christeson, G.~L., Barker, D.~H., Austin, J.~A., \& Dalziel, I.~W., 2003.
\newblock Deep crustal structure of bransfield strait: Initiation of a back arc
  basin by rift reactivation and propagation, {\it Journal of Geophysical
  Research: Solid Earth\/}, {\bf 108}(B10).

\bibitem[Clarke et~al.(2011)Clarke, Zaccarelli, Shapiro, \&
  Brenguier]{clarke2011assessment}
Clarke, D., Zaccarelli, L., Shapiro, N., \& Brenguier, F., 2011.
\newblock Assessment of resolution and accuracy of the moving window cross
  spectral technique for monitoring crustal temporal variations using ambient
  seismic noise, {\it Geophysical Journal International\/}, {\bf
  \textbf{186}}(2), 867--882.

\bibitem[Daskalakis et~al.(2016)Daskalakis, Evangelidis, Garnier, Melis,
  Papanicolaou, \& Tsogka]{daskalakis2016robust}
Daskalakis, E., Evangelidis, C., Garnier, J., Melis, N., Papanicolaou, G., \&
  Tsogka, C., 2016.
\newblock Robust seismic velocity change estimation using ambient noise
  recordings, {\it Geophysical Journal International\/}, {\bf 205}(3),
  1926--1936.

\bibitem[Davy et~al.(2016)Davy, Barruol, Fontaine, \&
  Cordier]{davy2016analyses}
Davy, C., Barruol, G., Fontaine, F.~R., \& Cordier, E., 2016.
\newblock Analyses of extreme swell events on la r{\'e}union island from
  microseismic noise, {\it Geophysical Supplements to the Monthly Notices of
  the Royal Astronomical Society\/}, {\bf 207}(3), 1767--1782.

\bibitem[De~Ridder \& Biondi(2013)]{de2013daily}
De~Ridder, S. \& Biondi, B., 2013.
\newblock Daily reservoir-scale subsurface monitoring using ambient seismic
  noise, {\it Geophysical Research Letters\/}, {\bf \textbf{40}}(12),
  2969--2974.

\bibitem[Draganov et~al.(2009)Draganov, Campman, Thorbecke, Verdel, \&
  Wapenaar]{draganov2009reflection}
Draganov, D., Campman, X., Thorbecke, J., Verdel, A., \& Wapenaar, K., 2009.
\newblock Reflection images from ambient seismic noise, {\it Geophysics\/},
  {\bf \textbf{74}}(5), A63--A67.

\bibitem[Duputel et~al.(2009)Duputel, Ferrazzini, Brenguier, Shapiro, Campillo,
  \& Nercessian]{duputel2009real}
Duputel, Z., Ferrazzini, V., Brenguier, F., Shapiro, N., Campillo, M., \&
  Nercessian, A., 2009.
\newblock Real time monitoring of relative velocity changes using ambient
  seismic noise at the piton de la fournaise volcano (la r{\'e}union) from
  january 2006 to june 2007, {\it Journal of Volcanology and Geothermal
  Research\/}, {\bf \textbf{184}}(1), 164--173.

\bibitem[Dziewonski et~al.(1969)Dziewonski, Bloch, \&
  Landisman]{dziewonski1969technique}
Dziewonski, A., Bloch, S., \& Landisman, M., 1969.
\newblock A technique for the analysis of transient seismic signals, {\it
  Bulletin of the seismological Society of America\/}, {\bf \textbf{59}}(1),
  427--444.

\bibitem[Dziewonski \& Anderson(1981)]{dziewonski1981preliminary}
Dziewonski, A.~M. \& Anderson, D.~L., 1981.
\newblock Preliminary reference earth model, {\it Physics of the earth and
  planetary interiors\/}, {\bf 25}(4), 297--356.

\bibitem[Gong et~al.(2015)Gong, Shen, Li, Li, \& Jia]{gong2015effects}
Gong, M., Shen, Y., Li, H., Li, X., \& Jia, J., 2015.
\newblock Effects of seasonal changes in ambient noise sources on monitoring
  temporal variations in crustal properties, {\it Journal of Seismology\/},
  {\bf 19}(3), 781--790.

\bibitem[Grad et~al.(1997)Grad, Shiobara, Janik, Guterch, \&
  Shimamura]{grad1997crustal}
Grad, M., Shiobara, H., Janik, T., Guterch, A., \& Shimamura, H., 1997.
\newblock Crustal model of the bransfield rift, west antarctica, from detailed
  obs refraction experiments, {\it Geophysical Journal International\/}, {\bf
  130}(2), 506--518.

\bibitem[Grob et~al.(2011)Grob, Maggi, \& Stutzmann]{grob2011observations}
Grob, M., Maggi, A., \& Stutzmann, E., 2011.
\newblock Observations of the seasonality of the antarctic microseismic signal,
  and its association to sea ice variability, {\it Geophysical Research
  Letters\/}, {\bf 38}(11).

\bibitem[Hadziioannou et~al.(2009)Hadziioannou, Larose, Coutant, Roux, \&
  Campillo]{hadziioannou2009stability}
Hadziioannou, C., Larose, E., Coutant, O., Roux, P., \& Campillo, M., 2009.
\newblock Stability of monitoring weak changes in multiply scattering media
  with ambient noise correlation: Laboratory experiments, {\it The Journal of
  the Acoustical Society of America\/}, {\bf \textbf{125}}(6), 3688--3695.

\bibitem[Heeszel et~al.(2016)Heeszel, Wiens, Anandakrishnan, Aster, Dalziel,
  Huerta, Nyblade, Wilson, \& Winberry]{heeszel2016upper}
Heeszel, D.~S., Wiens, D.~A., Anandakrishnan, S., Aster, R.~C., Dalziel, I.~W.,
  Huerta, A.~D., Nyblade, A.~A., Wilson, T.~J., \& Winberry, J.~P., 2016.
\newblock Upper mantle structure of central and west antarctica from array
  analysis of rayleigh wave phase velocities, {\it Journal of Geophysical
  Research: Solid Earth\/}, {\bf 121}(3), 1758--1775.

\bibitem[Herrmann(2013)]{herrmann2013computer}
Herrmann, R.~B., 2013.
\newblock Computer programs in seismology: An evolving tool for instruction and
  research, {\it Seismological Research Letters\/}, {\bf 84}(6), 1081--1088.

\bibitem[Janik(1997)]{janik1997seismic}
Janik, T., 1997.
\newblock Seismic crustal structure of the bransfield strait, west antarctica,
  {\it Polish polar research\/}, {\bf 18}(3-4), 171--225.

\bibitem[Janik et~al.(2006)Janik, {\'S}roda, Grad, \& Guterch]{janik2006moho}
Janik, T., {\'S}roda, P., Grad, M., \& Guterch, A., 2006.
\newblock Moho depth along the antarctic peninsula and crustal structure across
  the landward projection of the hero fracture zone, in {\em Antarctica\/}, pp.
  229--236, Springer.

\bibitem[Lepore \& Grad(2018)]{lepore2018analysis}
Lepore, S. \& Grad, M., 2018.
\newblock Analysis of the primary and secondary microseisms in the wavefield of
  the ambient noise recorded in northern poland, {\it Acta Geophysica\/}, {\bf
  66}(5), 915--929.

\bibitem[Levshin et~al.(1989)Levshin, Yanovskaya, Lander, Bukchin, Barmin,
  Ratnikova, \& Its]{levshin1989seismic}
Levshin, A., Yanovskaya, T., Lander, A., Bukchin, B., Barmin, M., Ratnikova,
  L., \& Its, E., 1989.
\newblock Seismic surface waves in a laterally inhomogeneous earth, {\it Modern
  Approaches in Geophysics\/}, {\bf \textbf{9}}, 131--169.

\bibitem[Mainsant et~al.(2012)Mainsant, Larose, Br{\"o}nnimann, Jongmans,
  Michoud, \& Jaboyedoff]{mainsant2012ambient}
Mainsant, G., Larose, E., Br{\"o}nnimann, C., Jongmans, D., Michoud, C., \&
  Jaboyedoff, M., 2012.
\newblock Ambient seismic noise monitoring of a clay landslide: Toward failure
  prediction, {\it Journal of Geophysical Research: Earth Surface\/}, {\bf
  \textbf{117}}(F1).

\bibitem[Meier et~al.(2010)Meier, Shapiro, \& Brenguier]{meier2010detecting}
Meier, U., Shapiro, N.~M., \& Brenguier, F., 2010.
\newblock Detecting seasonal variations in seismic velocities within los
  angeles basin from correlations of ambient seismic noise, {\it Geophysical
  Journal International\/}, {\bf \textbf{181}}(2), 985--996.

\bibitem[Mordret et~al.(2016)Mordret, Mikesell, Harig, Lipovsky, \&
  Prieto]{mordret2016monitoring}
Mordret, A., Mikesell, T.~D., Harig, C., Lipovsky, B.~P., \& Prieto, G.~A.,
  2016.
\newblock Monitoring southwest greenland’s ice sheet melt with ambient
  seismic noise, {\it Science Advances\/}, {\bf \textbf{2}}(5), e1501538.

\bibitem[Nicolson et~al.(2012)Nicolson, Curtis, Baptie, \&
  Galetti]{nicolson2012seismic}
Nicolson, H., Curtis, A., Baptie, B., \& Galetti, E., 2012.
\newblock Seismic interferometry and ambient noise tomography in the british
  isles, {\it Proceedings of the Geologists' Association\/}, {\bf
  \textbf{1}}(1), 74--86.

\bibitem[Obermann et~al.(2015)Obermann, Kraft, Larose, \&
  Wiemer]{obermann2015potential}
Obermann, A., Kraft, T., Larose, E., \& Wiemer, S., 2015.
\newblock Potential of ambient seismic noise techniques to monitor the st.
  gallen geothermal site (switzerland), {\it Journal of Geophysical Research:
  Solid Earth\/}, {\bf \textbf{120}}(6), 4301--4316.

\bibitem[Obermann et~al.(2016)Obermann, Lupi, Mordret, Jakobsd{\'o}ttir, \&
  Miller]{obermann20163d}
Obermann, A., Lupi, M., Mordret, A., Jakobsd{\'o}ttir, S.~S., \& Miller, S.~A.,
  2016.
\newblock 3d-ambient noise rayleigh wave tomography of sn{\ae}fellsj{\"o}kull
  volcano, iceland, {\it Journal of Volcanology and Geothermal Research\/},
  {\bf \textbf{317}}, 42--52.

\bibitem[Park et~al.(2012)Park, Kim, Lee, Yoo, \& Plasencia~L]{park2012p}
Park, Y., Kim, K.-H., Lee, J., Yoo, H.~J., \& Plasencia~L, M.~P., 2012.
\newblock P-wave velocity structure beneath the northern antarctic peninsula:
  evidence of a steeply subducting slab and a deep-rooted low-velocity anomaly
  beneath the central bransfield basin, {\it Geophysical Journal
  International\/}, {\bf 191}(3), 932--938.

\bibitem[Pratt et~al.(2017)Pratt, Wiens, Winberry, Anandakrishnan, \&
  Euler]{pratt2017implications}
Pratt, M.~J., Wiens, D.~A., Winberry, J.~P., Anandakrishnan, S., \& Euler,
  G.~G., 2017.
\newblock Implications of sea ice on southern ocean microseisms detected by a
  seismic array in west antarctica, {\it Geophysical Journal International\/},
  {\bf 209}(1), 492--507.

\bibitem[Quiros et~al.(2016)Quiros, Brown, \& Kim]{quiros2016seismic}
Quiros, D.~A., Brown, L.~D., \& Kim, D., 2016.
\newblock Seismic interferometry of railroad induced ground motions: body and
  surface wave imaging, {\it Geophysical Journal International\/}, {\bf
  \textbf{205}}(1), 301--313.

\bibitem[Ritzwoller et~al.(2001)Ritzwoller, Shapiro, Levshin, \&
  Leahy]{ritzwoller2001crustal}
Ritzwoller, M.~H., Shapiro, N.~M., Levshin, A.~L., \& Leahy, G.~M., 2001.
\newblock Crustal and upper mantle structure beneath antarctica and surrounding
  oceans, {\it Journal of Geophysical Research: Solid Earth\/}, {\bf 106}(B12),
  30645--30670.

\bibitem[Ryberg(2011)]{ryberg2011body}
Ryberg, T., 2011.
\newblock Body wave observations from cross-correlations of ambient seismic
  noise: A case study from the karoo, rsa, {\it Geophysical Research
  Letters\/}, {\bf \textbf{38}}(13).

\bibitem[Sabra et~al.(2005)Sabra, Gerstoft, Roux, Kuperman, \&
  Fehler]{sabra2005extracting}
Sabra, K.~G., Gerstoft, P., Roux, P., Kuperman, W., \& Fehler, M.~C., 2005.
\newblock Extracting time-domain green's function estimates from ambient
  seismic noise, {\it Geophysical Research Letters\/}, {\bf \textbf{32}}(3).

\bibitem[Sens-Sch{\"o}nfelder \& Wegler(2006)]{sens2006passive}
Sens-Sch{\"o}nfelder, C. \& Wegler, U., 2006.
\newblock Passive image interferometry and seasonal variations of seismic
  velocities at merapi volcano, indonesia, {\it Geophysical research
  letters\/}, {\bf \textbf{33}}(21).

\bibitem[Shapiro \& Campillo(2004)]{shapiro2004emergence}
Shapiro, N.~M. \& Campillo, M., 2004.
\newblock Emergence of broadband rayleigh waves from correlations of the
  ambient seismic noise, {\it Geophysical Research Letters\/}, {\bf
  \textbf{31}}(7).

\bibitem[Shapiro et~al.(2005)Shapiro, Campillo, Stehly, \&
  Ritzwoller]{shapiro2005high}
Shapiro, N.~M., Campillo, M., Stehly, L., \& Ritzwoller, M.~H., 2005.
\newblock High-resolution surface-wave tomography from ambient seismic noise,
  {\it Science\/}, {\bf \textbf{307}}(5715), 1615--1618.

\bibitem[Toyokuni et~al.(2018)Toyokuni, Takenaka, Takagi, Kanao, Tsuboi, Tono,
  Childs, \& Zhao]{toyokuni2018changes}
Toyokuni, G., Takenaka, H., Takagi, R., Kanao, M., Tsuboi, S., Tono, Y.,
  Childs, D., \& Zhao, D., 2018.
\newblock Changes in greenland ice bed conditions inferred from seismology,
  {\it Physics of the Earth and Planetary Interiors\/}, {\bf 277}, 81--98.

\bibitem[Vuan et~al.(2005)Vuan, Maurice, Wiens, \& Panza]{vuan2005crustal}
Vuan, A., Maurice, S.~R., Wiens, D., \& Panza, G., 2005.
\newblock Crustal and upper mantle s-wave velocity structure beneath the
  bransfield strait (west antarctica) from regional surface wave tomography,
  {\it Tectonophysics\/}, {\bf 397}(3-4), 241--259.

\bibitem[Vuan et~al.(2014)Vuan, Sugan, \& Linares]{vuan2014reappraisal}
Vuan, A., Sugan, M., \& Linares, M.~P., 2014.
\newblock A reappraisal of surface wave group velocity tomography in the
  subantarctic scotia sea and surrounding ridges, {\it Global and Planetary
  Change\/}, {\bf 123}, 223--238.

\bibitem[Weaver et~al.(2011)Weaver, Hadziioannou, Larose, \&
  Campillo]{weaver2011precision}
Weaver, R.~L., Hadziioannou, C., Larose, E., \& Campillo, M., 2011.
\newblock On the precision of noise correlation interferometry, {\it
  Geophysical Journal International\/}, {\bf \textbf{185}}(3), 1384--1392.

\bibitem[Wegler \& Sens-Sch{\"o}nfelder(2007)]{wegler2007fault}
Wegler, U. \& Sens-Sch{\"o}nfelder, C., 2007.
\newblock Fault zone monitoring with passive image interferometry, {\it
  Geophysical Journal International\/}, {\bf \textbf{168}}(3), 1029--1033.

\bibitem[Yang et~al.(2007)Yang, Ritzwoller, Levshin, \&
  Shapiro]{yang2007ambient}
Yang, Y., Ritzwoller, M.~H., Levshin, A.~L., \& Shapiro, N.~M., 2007.
\newblock Ambient noise rayleigh wave tomography across europe, {\it
  Geophysical Journal International\/}, {\bf \textbf{168}}(1), 259--274.

\bibitem[Yegorova et~al.(2011)Yegorova, Bakhmutov, Janik, \&
  Grad]{yegorova2011joint}
Yegorova, T., Bakhmutov, V., Janik, T., \& Grad, M., 2011.
\newblock Joint geophysical and petrological models for the lithosphere
  structure of the antarctic peninsula continental margin, {\it Geophysical
  Journal International\/}, {\bf 184}(1), 90--110.

\end{thebibliography}

\label{lastpage}

\end{document}